\documentclass[sn-basic]{sn-jnl}

\usepackage{xcolor}
\usepackage{longtable}
\usepackage{amsfonts}
\usepackage{amsmath}
\usepackage{graphicx}
\usepackage[english]{babel}

\theoremstyle{thmstyleone}%
\theoremstyle{thmstyletwo}%

\theoremstyle{thmstylethree}%

\title[Is the Italian university ready to a gender leap? ]{Is the Italian university ready for a gender leap? A regional perspective}

\author*[1]{\fnm{Ilia} \sur{Negri}}\email{ilia.negri@unical.it}\equalcont{These authors contributed equally to this work.}

\author[2]{\fnm{Maura} \sur{Mezzetti}}\email{maura.mezzetti@uniroma2.it}
\equalcont{These authors contributed equally to this work.}

\affil[1]{\orgdiv{Department of Economics, Statistics and Finance}, \orgname{Universit\`{a} della Calabria}, \orgaddress{\street{Via P. Bucci}, \city{Rende (CS)}, \postcode{87036},  \country{Italy}}}

\affil[2]{\orgdiv{Department of Economics and Finance}, \orgname{Universit\`{a} ``Tor Vergata''}, \orgaddress{\street{Via Columbia 2}, \city{Rome}, \postcode{00133},  \country{Italy}}}

\date{}

\abstract{ A discussion on the readiness of Italian universities to address gender-related issues from a regional standpoint is proposed. A statistical analysis is conducted on data of all scholars enrolled in Italian universities from 2000 to 2023 to investigate why the glass ceiling of the full professor position remains so challenging to break in almost all scientific fields and across all regions of Italy. 
}



\begin{document}

\maketitle
\section{Introduction}

According to a recent legislation\footnote{Decree-law 36/2022} on the subject of equal
opportunities between men and women, public administrations (PA) must highlight
and always communicate the data with the gender dimension, i.e. differentiating the
information for both men and women. The Italian Department for Equal Opportunities has recently enacted a directive\footnote{The guidelines "Linee guida sulla parità di genere nell'organizzazione e gestione del rapporto di lavoro con le pubbliche amministrazioni” have been elaborated in line with the contents of article 5 of decree-law 36/2022 and were enacted on 06/10/2022} that appears to encounter significant challenges in implementation, particularly within universities, which, as special cases of public administrations and workplaces, are expected to demonstrate a heightened sensitivity towards equal opportunity. This expectation arises from the universities' role as exemplars within society.  In this paper, we examine the careers of all scholars in Italian universities from 2000 to 2023, paying special attention to the gender dimension. Specifically, we analyze the career trajectories of scholars enrolled in Italian universities during the new century, considering their career progression over the entire period and investigating whether the probability to be promoted to the highest levels is influenced by gender or other factors such as research area or region.
To account for all these factors, a logistic regression is applied to estimate the probability of promotion. While this idea is not new in this context \citep{beasley2006time, filandri2021, depaola2015competition}, the novelty of our work lies in the comprehensive analysis of the entire Italian university system to gain a deeper understanding of the overall situation.
In a recent study \citep{falco2023gender}, the entire dataset was considered, but they excluded all scholars with the same name. Here, we employ a new method of scholar identification (unfortunately, the available dataset lacks a unique identifier for each scholar), based on research area and, when applicable, university affiliation. Through this approach, our final dataset comprises over 109,000 observations compared to the 60,000 of \cite{falco2023gender}. 
The rest of the paper is organized as follows.
In the next Section \ref{data_description}, we describe the dataset and all the variables considered. In Section \ref{gender_dimension}, the gender dimension in Italian universities is presented, focusing on research areas and geographical regions.
In Section \ref{StatisticalModel}, the statistical model is presented along with the results, and finally, in Section \ref{Conclusion}, some conclusions and comments are presented.

\section{Data description}
\label{data_description}
We refer to data available on the web site of the Italian Ministry of University and Research (MUR) \cite{MURcerca_uni}. 
We have assembled a dataset covering all scholars in Italian universities from 2000 to 2023. We merged the downloaded data for each year and each area. Unfortunately we do not have a unique identification and name in not unique. After removing scholars with exactly the same name, university, and area, we obtained a total of 109,239 observations. We further eliminated almost 600 scholars, checking for career, region and areas. Moreover almost 1700 schoars in the Telematica Universities will not be inluded in the modelling the promotion.
For each scholar, we recorded their position and the university where they served at the end of each year. This information allowed us to reconstruct their careers, including all promotions, and track their mobility. Additionally, we have data on each scholar's gender, macro sector, and disciplinary scientific sector.
The number of scholars present in the dataset in 2000 and still present in 2023 is 14,692, so they were followed for the entire period. 

In the next subsection, we will clarify how university roles are classified in Italian universities, how the macro areas of research are categorized, how universities are classified, and how they are distributed throughout the Italian territory.

\subsection{Academic role classification}
\label{role}
In the last 30 years, Italian academic roles have seen a myriad of denominations and classifications emerge. Here, we have elaborated a classification that aligns with the European standard and encompasses all the diverse roles still prevalent in Italian universities. 
Grade A is the highest level in academia. In Italy, according to MIUR classifications, this corresponds to the role of ``Ordinario", also known as the Full Professor level, alternatively referred to as ``Professore di prima fascia"\footnote{Unfortunately, in Italian academia, the role corresponding to a full professor is only declined with the masculine gender as {\em Ordinario}. This leads to some misleading situations, such as a female full professor being referred to as {\em Professore Ordinario} (masculine) instead of {\em Professoressa Ordinaria}, as Italian grammar would require.}. Grade B corresponds to the position of Associate Professor or ``Professore di seconda fascia'' in Italian academia. Grade C corresponds to the position of ``Ricercatore RtdB". This position has existed since 2010, and after three years, scholars become Associate Professors, i.e., Grade B, if they obtain the corresponding qualification\footnote{From 2024, RtdB and RtdA no longer exist and are replaced by the \emph{Research Tenure Track (RTT)} position.}. Grade D is introduced to account for the ``Ricercatore RtdA'' position, which is a tenure track position. This position also existed from 2010, but it does not have an automatic promotion to the upper level, so after three or five years, if you don't get another position such as RtdB, you no longer belong to academia. In Italian academia, there is also a position called  "Ricercatori universitari a tempo indeterminato", which can be translated as the \emph{tenured assistant professor} position. This role seems to exist only in a few European countries and is equivalent to a permanent position at Grade C, where tenured positions are not typically found. Grade E is used to describe this group of tenured assistant professors. Moreover, in Italian academia, there still exist some different roles due to temporary laws, which very often in Italy become permanent laws, where very few people are enrolled at those levels. They are classified within one of the previous grades according to the tasks and duties of the different positions.

\subsection{Classification of Academic Research Areas}
Italian academic research areas are divided into 14 macro areas. They are described in Table \ref{tab:aree}.
\begin{table}[h]
    \centering
    \caption{Code, name, number of SSDs, and total number of scholars for the 14 research areas. The SSDs M-EDF/01 and M-EDF/02, present in both area 06 and area 11, are counted only once in the overall total. Year 2023.}
    \begin{tabular}{r|l|r | r |r }
    \hline
Code & Name &  SSD & Scholars  & \% women \\
\hline
01& Mathematics and informatics & 10 &3601 & 29.63\\
02& Physics & 8 & 2792  & 23.39 \\ 
03& Chemistry & 12& 3312 & 50\\
04& Earth sciences & 12 &1192 & 29.45\\
05& Biology & 19 & 5402 & 56.03\\
06& Medicine & 52 & 9682 & 38.19\\
07& Agricultural and veterinary sciences &30& 3439 & 42.83 \\
08& Civil engineering and architecture & 25& 4106 & 38.41\\
09& Industrial and information engineering & 39& 7231 & 20.62\\
10& Antiquities, philology, literary st., art history & 77& 5219 & 54.56\\
11& History, philosophy, pedagogy and psychology & 34& 4998 & 49.57\\
12& Law & 21& 4998 & 40.20\\
13& Economics and statistics & 19& 5769 & 39.76\\
14& Political and social sciences & 14&  1977 & 41.83\\
\hline 
Tot&& 370& 63703 & 39.94\\
\hline 
\end{tabular}
         
    \label{tab:aree}
\end{table}
Every area contains different scientific disciplinary sectors, referred to in Italian as 
``Settore Scientifico Disciplinare'' (SSD).  There is a total of 370 SSD whose number vary in different disciplinary area.  Although STEM (Science, Technology, Engineering, and Mathematics) disciplines are related to the classification of degree courses, we can consider Areas 1 to 4 and Area 9 as STEM areas. In Table \ref{tab:aree} we report also the total number of scholar in each macro areas at the end of 2023, and for the same period the percentage of women. 
We notice a large heterogeneity in the number of scholars across macro-areas. In only three sectors (05 Biology, and 10 Antiquities, philology, literary studies, and art history) is the percentage of women equal to or greater than 50\%, while in Sector 11 (History, philosophy, pedagogy, and psychology), it almost reaches 50\%. Four macro-areas of research have a percentage of women less than 30\%, with the lowest level reached by sector 09 (Industrial and information engineering), with 20.62\% of women. 

\subsection{Geographical distribution of Italian universities } 
In Italy in 2023, there were 98 universities. The names of these universities, listed in alphabetical order as recorded in the MUR database, along with their types (the majority being national, 17 private, and 11 online), and their respective geographical regions are reported in Table \ref{region}. In Table \ref{tab:region}, we report the number of universities and the number of scholars for each of the 20 political regions in Italy, along with the percentage of women. For further details on this indicator, please refer to the bottom panels of Figure \ref{fig:woman_total_full}. The percentage of women ranges from 31\% in Trentino AA (North) to 44\% in Valle d'Aosta (also located in the North). Moreover, only the 12\% of the universities have a female rector (or dean). This is the first indicator of how difficult it is to break the glass ceiling in Italian universities. Specifically, only 12 universities in Italy are led by a female leader. In terms of territorial distribution, the majority of universities, totaling 33, are located in the North of Italy, with 14 of them situated in Lombardy. In the central part of Italy, there are 31 universities, while in the South, there are 23 universities. If the number of universities can be considered as an indicator of regional educational development or institutional density \citep{guironnet2018geographical},  these figures underscore the substantial variation between the North and South of Italy.

\begin{table}[h]
    \centering
    \caption{Name, number of universities and number of scholar in each of the 20 Italian political region. Year 2023.}
    \begin{tabular}{l|r | r | r }
    \hline
Region & Univeristy & Scholars & Female \%  \\
\hline
PIEMONTE	&   4 &4026 & 40.19 \\ 
VALLE D’AOSTA   &1 & 56 & 44.64 \\ 
LOMBARDIA		& 14 & 10395 &  39.15 \\
TRENTINO-ALTO ADIGE &2 & 1113 & 31.09 \\
VENETO			&4 & 4410 & 39.61 \\ 
FRIULI-VENEZIA GIULIA &3 & 1492 &33.24 \\
LIGURIA			&1 & 1408 & 39.56 \\ 
EMILIA ROMAGNA	& 4 & 6059 & 40.88 \\
TOSCANA			& 7 & 4786 & 39.01 \\ 
UMBRIA			& 2 & 1040 & 40.86 \\ 
MARCHE			& 4 & 1629 & 42.26 \\
LAZIO			& 13 & 7282 & 39.54 \\ 
ABRUZZO			& 4 & 1726 & 40.85 \\
MOLISE			& 1 & 321 & 34.58 \\ 
CAMPANIA		& 7 & 6059 & 42.65 \\ 
PUGLIA			& 5 & 3083 & 41.32 \\ 
BASILICATA		&1 & 323 & 38.08 \\
CALABRIA		& 4 & 1495 & 39.93 \\
SICILIA			& 4 & 4396 & 41.40 \\
SARDEGNA		& 2 & 1755 & 39.03 \\ 
\hline
ONLINE          & 11 &849&40.75\\  
\hline 
Total & 98 & 63703 & { 39.94}\\
\end{tabular}
    \label{tab:region}
\end{table}

\section{Gender dimension in Italian Universities}
\label{gender_dimension}

Let us give some idea on gender dimension in Italian universities with a focus on the highest position (Grade A). The analysis is done along four different directions: role, research area, geographical region and promotion to highest role. The dynamics between these different aspects and the impact of the first three on the latest one will be analyzed through a statistical model 
in Section \ref{StatisticalModel}.  
\begin{table}[h]
    \centering
    \caption{Percentage of woman in Italian universities according to the academic role in 2000}
    \begin{tabular}{l|rrr}
       Role & Female  & Female \%  & Total  \\
       \hline
  Grade A & 1992 & 13.41\% &   14850 \\
    Grade B &  4759 & 27.87\% &   17075\\
    Grade E & 8551  & 41.29\% &   20711\\
    \hline
    Total& 15302 & 29.10\% & 52636\\
    \end{tabular}
    
    \label{tab:ruolo_totale2000}
\end{table}
\subsection{Role dimension}
At the end of 2023 there were 63703 scholars and 25242 were women  (39.94\%). Table \ref{tab:ruolo_totale2023} describe the percentage of women in different role in 2023 in Italy. What is evident from Table \ref{tab:ruolo_totale2023} is the drastic jump in the percentage of women who hold Grade A. If for the lower grades the percentage is between 42\% and 50\%, in Grade A it becomes less than 28\%. Note that women are nearly at parity only in Grade E, corresponding to the lowest level, with practically zero probability of advancing in the role after 2023. 
\begin{table}[h]
    \centering
    \caption{Percentage of woman in Italian universities according to the academic role in 2023}
    \begin{tabular}{l|rrr}
       Role & Female  & Female \%  & Total  \\
       \hline
   Grade A &   4671 & 27.91\% & 16736 \\ 
 Grade B &11255 & 42.85\% & 26264 \\ 
 Grade C & 2857 & 41.87\% & 6824 \\ 
 Grade D &4267 & 46.97\% & 9085 \\ 
 Grade E &2392 & 49.89\% & 4794 \\ 
    \hline
    Total&  25442  &39.94\%&     63703\\
    \end{tabular}
        \label{tab:ruolo_totale2023}
\end{table}

If we refer to Table \ref{tab:ruolo_totale2000}, we observe that the situation has only seen a slight improvement over the past 20 years. In 2000, only 29\% of Italian scholars were women, but this percentage varies considerably across different roles. At Grade E, it stands at 41\%, then decreases to 28\% for Grade C, and further to 13\% for Grade A. Therefore, from 2000 to 2023, the situation has seen a modest improvement, with the percentage of women in Grade A rising from 13.4\% to 27.9\%. As explained earlier in Section \ref{role}, Grade B and Grade C did not exist in Italian universities in 2000.
\begin{table}[h]
    \centering
    \caption{Percentage distribution of role with respect to gender in Italian universities in 2000. }
    \label{teb:ruolo_percentuale2000}
    \begin{tabular}{l|rrr}
       Role & Female \% & Male \%  & Total  \% \\
       \hline
    Grade A & 13.2\% & 34.44\% &   28.321\% \\
    Grade B &  31.1\% & 32.99\% &   32.44\%\\
    Grade E & 55.88\%  & 32.56\% &   39.35\%\\
    \hline
    Total& 100\% & 100\% & 100\%\\
    \end{tabular}   
\end{table}
\begin{table}[h]
\centering
{   \caption{Percentage of woman in Italian universities according to the academic role in 2023}
 \label{tab:ruolo_percentuale2021}
    \begin{tabular}{l|rrr}
       Role & Female \% & Male \%  & Total  \% \\
       \hline
 Grade A &18.35\% & 31.53\% & 26.27\% \\ 
Grade B &44.24\% & 39.23\%& 41.23\% \\ 
Grade C &11.23\% & 10.37\% & 10.71\% \\ 
Grade D & 16.77\% & 12.59\% & 14.26\% \\ 
Grade E & 9.4\% & 6.28\%& 7.53\% \\ 
    \hline
    Total& 100\% & 100\% & 100\%\\
    \end{tabular}
}       
\end{table}
Another interesting insight into the gender difference in roles is provided by Table \ref{teb:ruolo_percentuale2000} and Table \ref{tab:ruolo_percentuale2021}  where the reported percentage in the first line can be read as penetration rate for Grade A position. In 2000, the distribution of roles among men is almost uniform, with each role having a similar representation. Conversely, among women, it is observed that over 50\% occupy the lowest grade, while only 13\% are able to attain the highest role, Grade A.

In {2023}, the situation becomes slightly less clear with the introduction of five roles instead of three. However, Table \ref{tab:ruolo_percentuale2021} reveals a significant gender disparity, notably in the penetration rate for Grade A positions, which stands at only 18\% for women compared to 31\% for men.

\subsection{Research area dimension}
The analysis of the distribution by gender in Grade A across the 14 macro-areas of research is shown in Table \ref{tab:genere_area}.
In 2023, we observe that in the STEM macro-areas of research, the percentage of women in Grade A remains consistently low, with the lowest percentages being 16\% and 14\% in Areas 2 and 9 respectively, despite continuous growth since 2000. But what is truly astonishing is that in no area of top-level research do women comprise 50\%, even in fields like (5 and 10) where women are the majority.

\begin{table}[ht]
\centering
\caption{Proportion of female and male and ratio women to men by Macro Area for Grade A position in 2000 and 2023.}
\begin{tabular}{r|rr|r|rr|r}
&\multicolumn{3}{c|}{2000} & \multicolumn{3}{c}{2023}\\
  \hline
Area & F & M &Ratio& F & M& Ratio \\ 
  \hline
1 & 0.15 & 0.85&0.17 & 0.21 & 0.79&0.26 \\ 
  2 & 0.05 & 0.95 &0.05& 0.16 & 0.84&0.19 \\ 
  3 & 0.11 & 0.89 & 0.12&0.37 & 0.63&0.59 \\ 
  4 & 0.09 & 0.91 & 0.10&0.20 & 0.80&0.25 \\ 
  5 & 0.22 & 0.78 & 0.29&0.39 & 0.61&0.65 \\ 
  6 & 0.07 & 0.93 &0.07& 0.22 & 0.78&0.28 \\ 
  7 & 0.09 & 0.91 &0.10& 0.26 & 0.74&0.36 \\
  8 & 0.09 & 0.91 & 0.10&0.27 & 0.73&0.36 \\ 
  9 & 0.03 & 0.97 &0.03& 0.14 & 0.86&0.16 \\ 
  10 & 0.33 & 0.67 &0.50& 0.46 & 0.54&0.86 \\ 
  11 & 0.22 & 0.78&0.28 & 0.42 & 0.58&0.71 \\ 
  12 & 0.11 & 0.89 & 0.12&0.29 & 0.71&0.41 \\ 
  13 & 0.12 & 0.88 &0.14& 0.27 & 0.73&0.38 \\ 
  14 & 0.15 & 0.85&0.17 & 0.31 & 0.69&0.45 \\ 
  \hline
  Total & 0.13 & 0.87&0.16& 0.28& 0.72&0.37  \\
   \hline
\end{tabular}
    \label{tab:genere_area}
\end{table}

Another indicator of gender disparity is provided by the femininity ratio, which represents the ratio of women to men. This ratio is reported for the years 2000 and {2023} across the 14 different macro-areas in Table \ref{tab:genere_area}.
Even today, at the highest level, there are only 37 women for every 100 men.
Although the situation has steadily improved over the last 20 years, with a gradual increase from 2000 to 2023 (where there were 16 women in Grade A for every 100 men), it still falls short of achieving gender parity. Many differences persist across the various macro-areas.
For example, in areas 2 (Physics) and 9 (Industrial and information engineering), the gender disparity is most pronounced, with only 18 and 15 women for every 100 men, respectively. In macro-area 10, women outnumber men, but they still fall short of achieving gender parity.


\subsection{Geographical dimension}
We now examine the gender dimension across the 20 regions of Italy. The economic, social, and cultural differences between the North and South of Italy are well-known and documented in various studies, see for example  \cite{aiello2000uneven}, \cite{andreotti2013female}, \cite{aiello2012structural}, and references therein. Of particular interest, and as revealed by our study, is the nearly complete nullification of these differences when considering the status of women in top university positions.

\begin{figure}
\centering
\includegraphics[width=\textwidth]{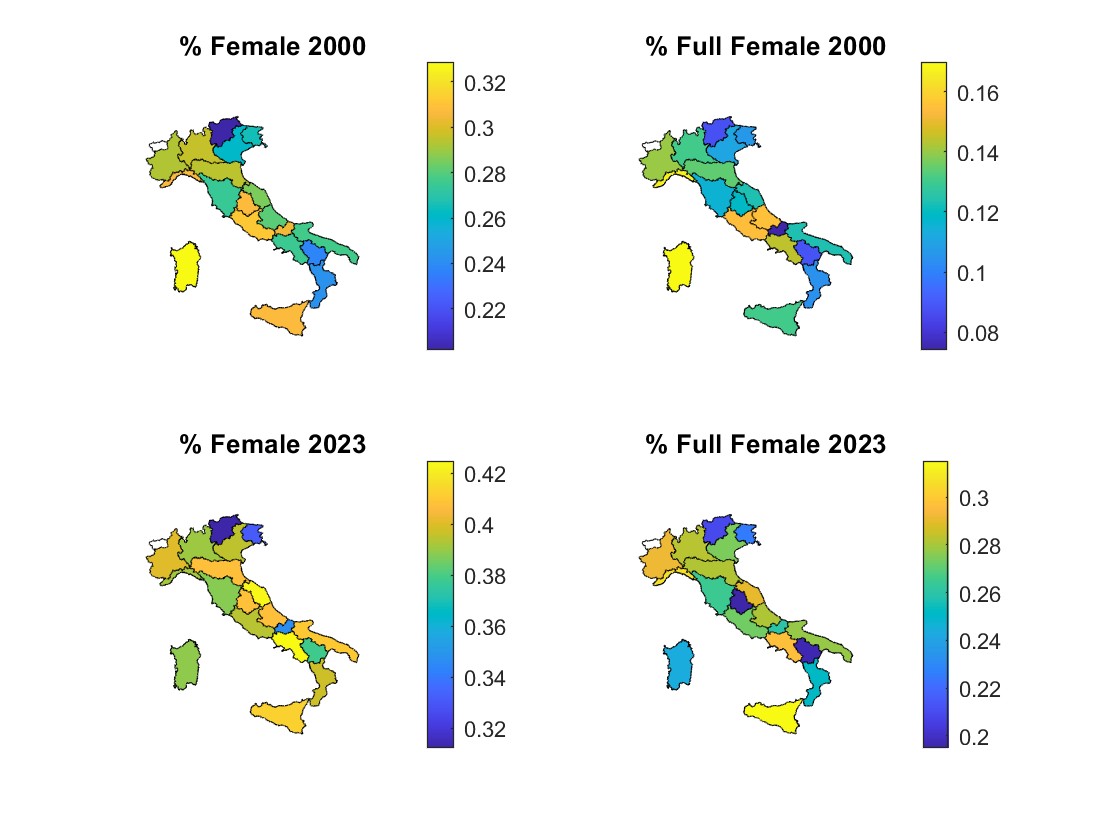}
\caption{Percentage of women and woman in Grade A position in Italian universities, in 2000 and in 2023}
\label{fig:woman_total_full}
\end{figure}

\begin{figure}
\centering
\includegraphics[width=\textwidth]{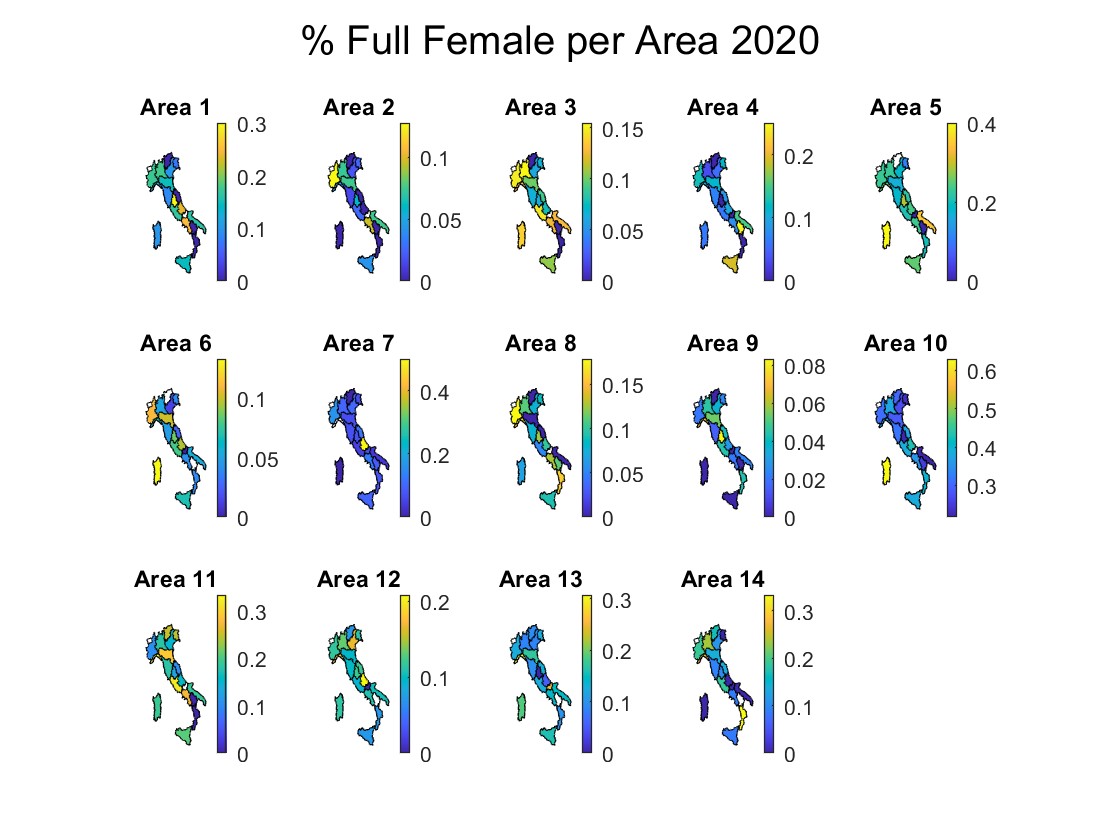}
\caption{Percentage of women in Grade A position in Italian university in each Macro Area, in 2000.} 
\label{fig:woman_full_area_2000}
\end{figure}

\begin{figure}
\centering
\includegraphics[width=\textwidth]{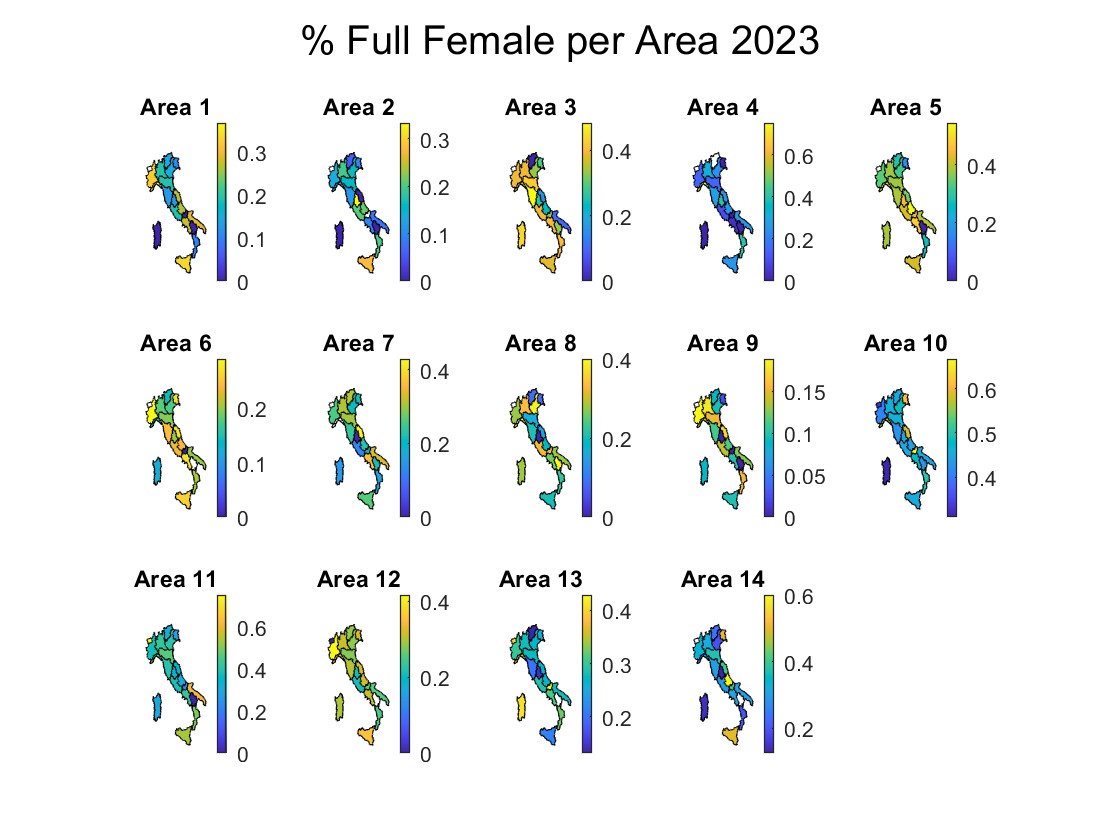}
\caption{Percentage of women in Grade A position in Italian university in each Macro Area, in 2023.} 
\label{fig:woman_full_area_2020}
\end{figure}

Indeed, as illustrated in Figure \ref{fig:woman_total_full}, the proportion of women in universities approaches 50\%, yet drops below 30\% across nearly all regions when considering those occupying Grade A positions. While there has been a slight improvement in all regions by 2023, the percentage of Grade A scholars still remains significantly below 50\%.
An analysis of the distribution of women per area and per political region is shown in Figures \ref{fig:woman_full_area_2000} and \ref{fig:woman_full_area_2020}. In nearly all regions and across almost all areas, the majority of Grade A positions are held predominantly by men. 
The only sector where the presence of women exceeds 50\% in some regions is Area 10 (Antiquities, philology, literary studies, art history). The situation has improved slightly in some areas (5, 10, and 11) over the past 20 years, as depicted in Figure \ref{fig:woman_full_area_2020}, but in others, especially STEM fields such as 1, 2, and 9, the scarcity of women in the highest positions in academia remains very significant.

\subsection{Promotion dimension}

To understand the underrepresentation of women in senior positions within academia, we investigate the promotion dimension to Grade A in Italian universities. We analyze data related to promotion to Grade A position from Grade B and promotion to grade B from grade E. Other type of promotions are not considered because are very peculiar and don't affect the final model. Although there are some cases of direct promotion from Grade D to Grade B or A, they are very few and not important for understanding the phenomenon. Moreover, promotion from grade C to grade B is practically automatic after three years and only a few cases do not obtain this promotion. In Section \ref{StatisticalModel} we will use this data to estimate the probability of being promoted to Grade A and to Grade B, respectively.  Table \ref{promotionB-A} reports  the numbers of promotions from grade B to grade A, the percentage of women among the promoted individuals, and the percentage of women in Grade A, from 2000 to 2023. From the data reported in the table, it is observed that in years with fewer promotions due to the suspension of competitions caused by financial difficulties or government actions, the percentage of promoted women decreases.

Although the number of positions each year varies greatly, going from a minimum of 29 in 2009 to a maximum of 2151 in 2001, the percentage of women promoted is around 30\% with a minimum of 21\% to a maximum of 37\%. 
\begin{table}
\centering
\caption{Promoted to Grade A from Grade B. Years 2001 -- 2023}
\label{promotionB-A}
\begin{tabular}{r|rrrr}
  \hline
Year & Promoted Grade A & F. Promoted & \% F. Promoted &\% F. in Grade A\\ 
  \hline
  2001 & 2151 & 484 & 22.50 & 14.70 \\ 
2002 & 1583 & 411 & 25.96 & 15.69 \\ 
2003 & 352 & 89 & 25.28 & 16.04 \\ 
2004 & 571 & 153 & 26.79 & 16.49 \\ 
2005 & 1601 & 442 & 27.61 & 17.55 \\ 
2006 & 1079 & 303 & 28.08& 18.19 \\ 
2007 & 361 & 118 & 32.69 & 18.61 \\ 
2008 & 146 & 42 & 28.77 & 18.93 \\ 
2009 & 29 & 6 & 20.69 & 19.23 \\ 
2010 & 289 & 81 & 28.03 & 20.17 \\ 
2011 & 613 & 182 & 29.69 & 20.73 \\ 
2012 & 232 & 64 & 27.59 & 20.92 \\ 
2013 & 145 & 37 & 25.52 & 21.21 \\ 
2014 & 264 & 65 & 24.62 & 21.40 \\ 
2015 & 450 & 103 & 22.89 & 21.55 \\ 
2016 & 989 & 290 & 29.32 & 22.12 \\ 
2017 & 816 & 243 & 29.78 & 22.92 \\ 
2018 & 1240 & 387 & 31.21 & 23.69 \\ 
2019 & 1266 & 433 & 34.20 & 24.71 \\ 
2020 & 1174 & 363 & 30.93 & 25.29 \\ 
2021 & 1640 & 559 & 34.08 & 26.13 \\ 
2022 & 1148 & 421 & 36.67 & 26.90 \\ 
2023 & 1563 & 574 & 36.72 & 27.90 \\ 
   \hline
   Total & 19702 &5850 & 29.69\\
   \hline
\end{tabular}
\end{table}

\begin{table}
\caption{Promoted to Grade B from grade E. Years 2001 -- 2023}
\centering
\begin{tabular}{r|rrrr}
  \hline
Year & Promoted Grade B & F Promoted & \% F Promoted&\% F in Grade E \\
  \hline
2001 & 2825 & 1019 & 36.07 & 42.28 \\ 
2002 & 2202 & 781 & 35.47 & 43.02 \\ 
2003 & 399 & 165 & 41.35 & 43.10 \\ 
2004 & 821 & 280 & 34.10 & 43.54 \\ 
2005 & 2309 & 866 & 37.51 & 44.43 \\ 
2006 & 1390 & 550 & 39.57 & 45.01 \\ 
2007 & 425 & 166 & 39.07 & 45.04 \\ 
2008 & 183 & 72 & 39.34 & 45.03 \\ 
2009 & 28 & 13 & 46.43 & 45.19 \\ 
2010 & 665 & 238 & 35.79 & 45.31 \\ 
2011 & 955 & 365 & 38.22 & 45.35 \\ 
2012 & 347 & 132 & 38.04 & 45.57 \\ 
2013 & 170 & 57 & 33.53 & 45.69 \\ 
2014 & 2273 & 869 & 38.23 & 46.52 \\ 
2015 & 3101 & 1246 & 40.18 & 47.66 \\ 
2016 & 1121 & 470 & 41.93 & 47.99 \\ 
2017 & 1078 & 446 & 41.37 & 48.41 \\ 
2018 & 1601 & 694 & 43.35 & 49.27 \\ 
2019 & 1602 & 774 & 48.31 & 49.61 \\ 
2020 & 1364 & 686 & 50.29 & 49.70 \\ 
2021 & 1068 & 546 & 51.12 & 49.92 \\ 
2022 & 1951 & 1041 & 53.36 & 49.18 \\ 
2023 & 188 & 94 & 50.00 & 49.80 \\ 
   \hline
   Total &28066 &11570&41.22&\\
   \hline   
\end{tabular}
\end{table}

\section{Statistical Model}
\label{StatisticalModel}
In a first approach we want to understand if the probability to be promoted to the Grade A position is affected by gender bias. We first consider all the academics, and did not consider dependence on scientific productivity for two principal reason. The first is that there is a very huge amount of literature that consider the effect of productivity on promotion and it is well known the effect of the so called 
``productivity gap'' in favour of men on promotion. Nevertheless in almost all published works only some subgroups of population corresponding to specific SSD are taken into consideration because reconstructing productivity is very difficult due to the great variety of scientific sectors and the impossibility of defining a homogeneous indicator for productivity. For example, in \cite{mezzettinegri2024} productivity is measured with a new measure introduced in \cite{mezzettinegri2023}, and it is included in the model for two subgroups of population.

On the other hand, considering the systemic scarcity of women across various dimensions within the entire Italian university system, as depicted in the previous section, our aim is to check if  every researcher  have equal opportunities for promotion to higher levels given they have attained that academic level. The premise of this idea is that every researcher who has attained an academic level possesses the minimum qualifications required by the SSD and the university to apply for a higher position.

In order to understand how the probability of being promoted is affected by other factors, we consider a logistic regression model where the probability of promotion depends on gender, and we incorporate fixed effects for the macro-area of research, the geographical region, and other quantities such as the time elapsed before promotion.

\subsection{Logistic Regression Models}
First, we consider a model for promotion to Grade A from Grade B. In this model, the response variable is the promotion from Associate Professor to Full Professor. We consider only academics who have held the title of 
``Associate Professor'' at least once from 2000 to 2023, totaling nearly 60,000 professors. 

Let $Y_{ijhk}$ represent the response variable for individual $i$, that at year $k$, is a researcher in area $j$, belonging to a university in region $h$. $Y_{ijhk}=1$ if the researcher is promoted to Full Professor in year $k$, and $Y_{ijhk}=0$ if the researcher is still an Associate Professor at year $k$, which is the last time the researcher is observed. 
Each researcher is observed only once, either at the time of promotion (with $Y_{ijhk}=1$) or at their last observation.

The waiting time is calculated as the duration, in years, that a researcher remains in the position of Associate Professor (Grade B).
The waiting time for promotion results in a left and right censored variable. This is because we may not observe the entry time in academia, and the last observation may either be followed by a promotion at a higher level or may never be followed by a promotion. Additionally, we do not know for how much longer the individual will be observed.
To address the issue of left censoring, we excluded more than 30,000 individuals who held the position of Associate Professor in 2000, as we were unable to estimate their date of entry.

We model the promotion event with a logistic regression, where a function of the probability to be promoted, denoted as
$E(Y_{ijhk})$, is a linear function of gender, waiting time, and the squared waiting time. We include scientific area, region, and year of promotion or year of entry as fixed effects.
Online universities were excluded because the mechanisms for career advancement can differ significantly from those in traditional universities. Finally, the population consists of nearly 30,000 individuals, among whom 24.9\% experience promotion from Grade B to Grade A. The average waiting time for promotion is 7.59 years (7.40 years for males and 8.02 years for females); both mean and median values show significant differences.

The waiting time for promotion varies across different areas; the minimum average value is observed for males in Area 1, with a mean value of 6.20 years, while the maximum is observed for women in Area 10, with a mean value of 8.97 years. Additionally, Area 1 exhibits the highest difference in average waiting time between males and females. 
Surprisingly, in Area 2, where women are underrepresented, the average waiting time for women is 4 months less than that for male colleagues. Therefore, women are promoted less frequently, but they experience shorter waiting times for promotion.

The regions have been reclassified into four macro areas: (i) North East, (ii) North West, (iii) Center, and (iv) South and Islands.  Moreover, both the time of entry and the year of promotion (which are considered alternatively) are reclassified into 5 classes. 
The results of the logistic regression are presented in Table \ref{tab:logistic1}, where the year of entry is considered. 
Overall, the logistic regression results provide valuable insights into the factors influencing promotion from Grade B to Grade A.
First of all, the coefficient for the gender effect (female versus male) is -0.619, with a 95\% confidence interval of (-0.68, -0.56),  indicates that, on average, females are less likely to be promoted compared to males, holding other variables constant. Regarding waiting time, both the linear (0.317) and quadratic (-0.016) terms show significant associations with promotion. This suggests a non-linear relationship. Initially, as waiting time increases, the likelihood of promotion also increases, as indicated by the positive linear term. However, as waiting time continues to increase, the effect starts to diminish, as indicated by the negative quadratic term. This implies that while waiting time positively influences promotion likelihood, the effect weakens as waiting time grows, resulting in a diminishing rate of increase for promotion.  

\begin{figure}
    \centering
    \includegraphics[width=1\linewidth]{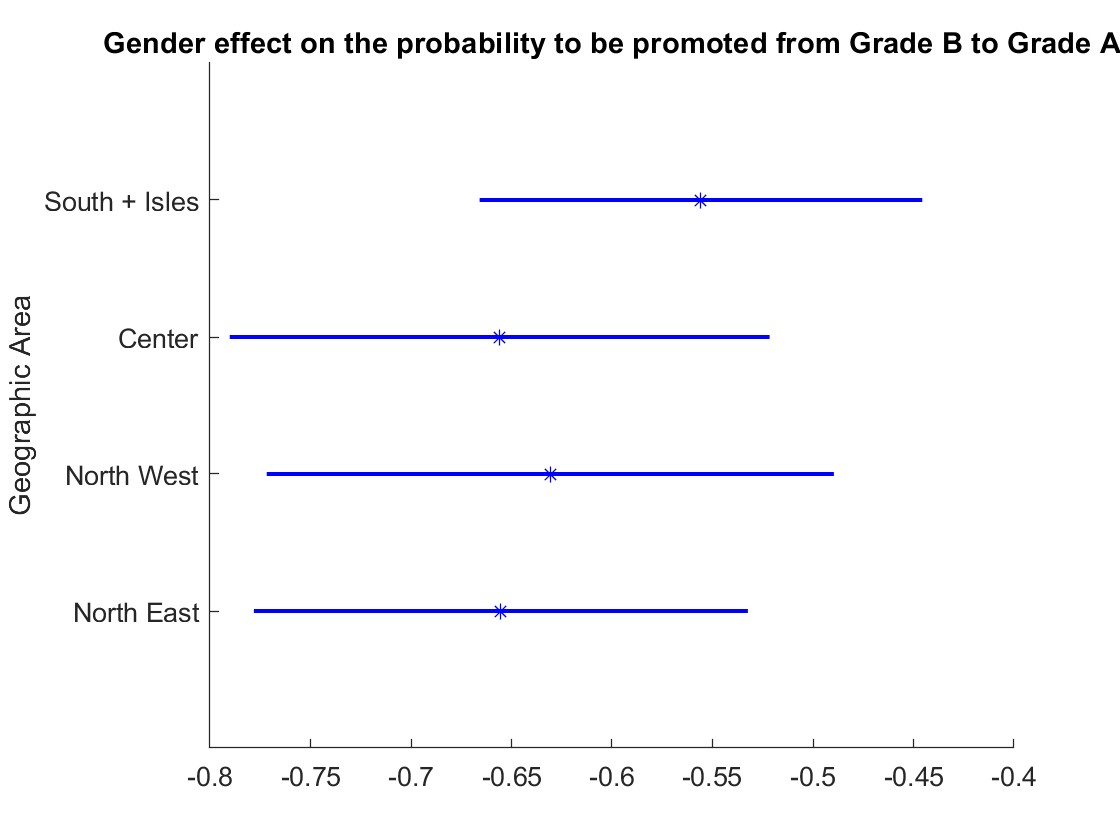}
    \caption{Coefficients for Probability to be be promoted (from Grade B to Grade A) for females versus males, stratified by geographic area}
\label{coefgender}
\end{figure}

\begin{figure}
    \centering
    \includegraphics[width=1\linewidth]{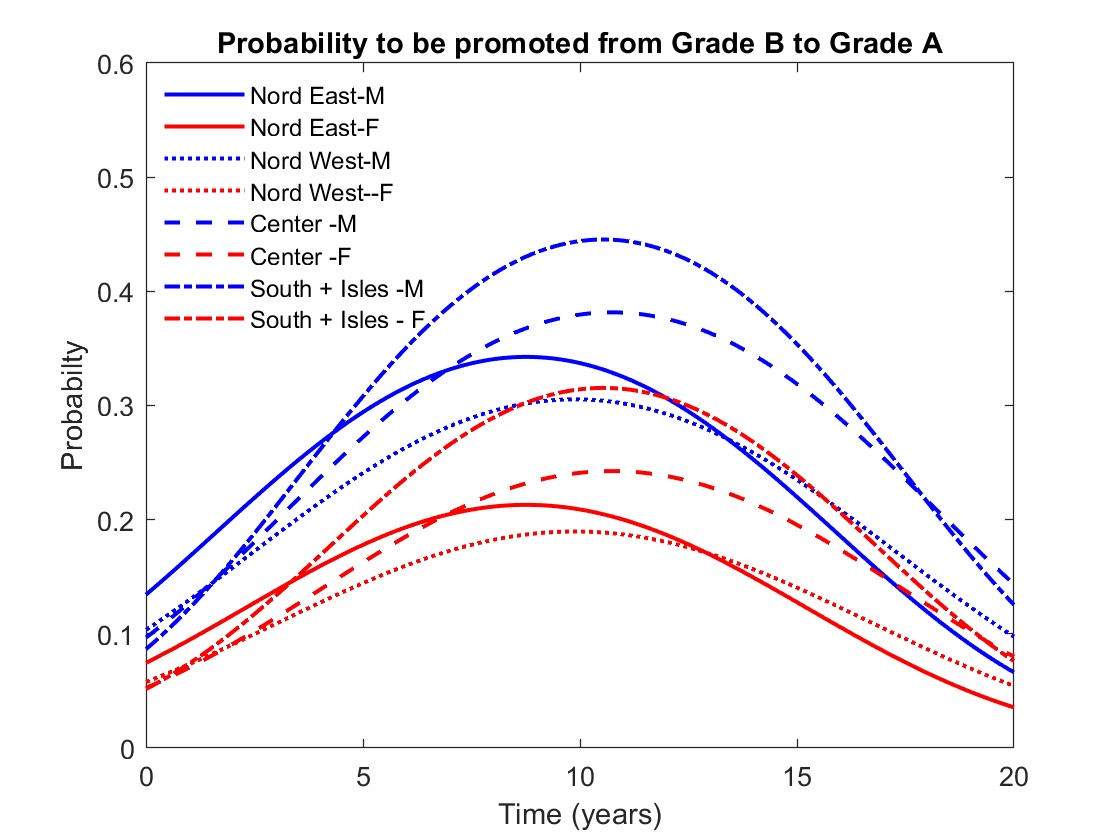}
    \caption{Probability to be promoted from Grade B to Grade A, as a function of waiting time for males and females, Stratified by geographic area}
\label{waitinggenderarea}
\end{figure}

To better understand the geographical heterogeneity in the promotion, we run stratified analysis by geographical area, including again as fixed effects the area and the period of entry.  In Figure \ref{coefgender}, the odds ratios (OR) and their corresponding $95\%$ confidence intervals are shown for the effect of gender (females versus males) when logistic regression is stratified for each area.
The relationship between the probability of promotion and waiting time is well described by a second-order relation, as shown in Figure \ref{waitinggenderarea}, where results from stratified analysis are shown. Once again, it is evident that the probability of promotion for a woman is consistently lower.

\begin{figure}
    \centering
    \includegraphics[width=1\linewidth]{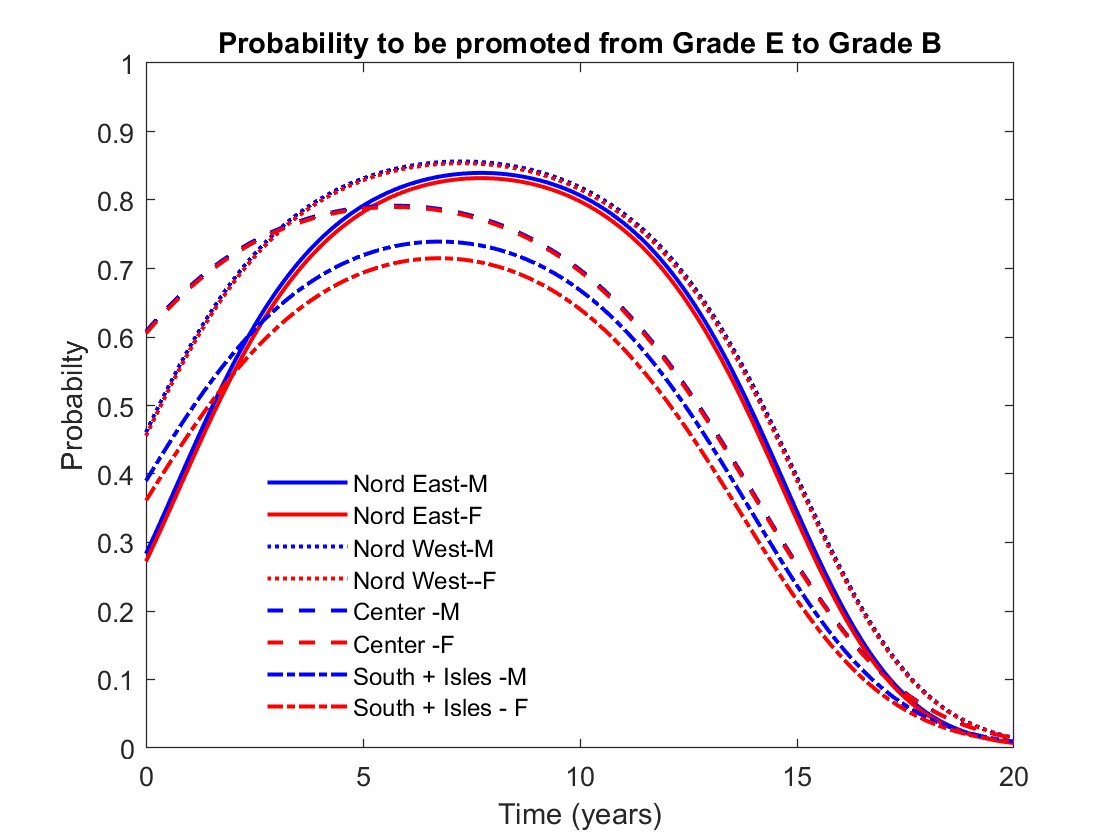}
    \caption{Probability to be promoted from Grade E to Grade B, as a function of waiting time for males and females, Stratified by geographic area}
\label{waitinggenderarea_etob}
\end{figure}

The same analysis has be carried out considering the promotion from B to grade E. 
As Grade E were positions that no longer exist, over the last two years there have been government incentives for their promotion, and during this time, 40\% of researchers have been promoted. Therefore, we have excluded promotions from the last two years to assess the gender effects of promotions.
In Table \ref{tab:logistic2} the results of the logistic with the same fixed effects. 
Stratifying the logistic regression,  only of South and Islands the effect of gender is significantly different from zero.  In Figure \ref{waitinggenderarea_etob} the geographical heterogeneity in  waiting time to promotion from Grade E to Grade B is shown. 



\begin{table}[]
    \centering
    \caption{Promotion from Grade B to Grade A: Estimation Results for a Logistic Model. Total sbservations: 29,360.}
    \label{tab:logistic1}
{
\begin{tabular}{l|rrrr}
  \hline
Variables & Estimate & Std. Error & z value & Pr($>|z|$) \\ 
\hline 
(Intercept) & -1.044 & 0.092 & -11.35 & $<0.001$ \\ 
gender& -0.619 & 0.032 & -19.46 &  $<0.001$ \\ 
waiting time & 0.317 & 0.012 & 25.59 &  $<0.001$  \\ 
waiting time$^2$ & -0.016 & 0.001 & -25.635 & $<0.001$  \\
\hline
NW vs NE & -0.186 & 0.04 & -4.243 &  $<0.001$  \\ 
Center vs NE & -0.275& 0.043& -6.42 & $<0.001$  \\ 
South + Is vs NE & -0.076 & 0.039 & -1.9383 & 0.05 \\ 
Area 2 vs 1 & -0.279 & 0.101 & -2.754 &  $<0.001$  \\ 
Area 3 vs 1  & 10.439 & 0.102 & -4.322 &  $<0.001$  \\ 
Area 4 vs 1  & -0.672 & 0.145 & -4.652 &  $<0.001$  \\ 
Area 5 vs 1  & -0.248 & 0.087 & -2.839 & 0.004 \\ 
Area 6 vs 1  & 0.103 & 0.076 & 1.363 & 0.17 \\ 
Area 7 vs 1  & -0.432 & 0.1 & -4.329 &  $<0.001$  \\ 
Area 8 vs 1  & -0.296 & 0.088 & -3.35 & $<0.001$ \\ 
Area 9 vs 1  & 0.002 & 0.081 & 0.027 & 0.98 \\ 
Area 10 vs 1  & -0.15 & 0.083 & -1.816 & 0.07 \\ 
Area 11 vs 1  & 0.109 & 0.081 & 1.335 & 0.18 \\ 
Area 12 vs 1  & 0.258 & 0.082 & 3.134 & 0.001 \\ 
 Area 13 vs 1 & 0.296 & 0.081 & 3.644 & $<0.001$ \\ 
Area 14 vs 1  & -0.044 & 0.101 & -0.439 & 0.66 \\ 
Year Entry 2005-10 vs $<2005$ & -0.638 & 0.035 & -18.04 & $<0.001$ \\ 
Year Entry 2010-15 vs $<2005$ & -1.263 & 0.0471 & -26.92 & $<0.001$ \\ 
Year Entry 2015-20 vs $<2005$ & -1.983 & 0.06 & -33.114 & $<0.001$ \\ 
Year Entry $>2020$ vs $<2005$ & -3.248 & 0.186 & -17.433 & $<0.001$ \\ 
\hline 
\end{tabular}
}
\end{table}

\begin{table}[]
    \centering
    \caption{Estimation result for a logistic model, Promotion from Grade E to Grade B. 21983 Observations.}
    \label{tab:logistic2}
{
\begin{tabular}{l|rrrr}
  \hline
Variables & Estimate & Std. Error & z value & Pr($>|z|$) \\ 
\hline 
(Intercept) & 2.264 & 0.141 & 16.051 & $<0.001$\\ 
gender & -0.066 & 0.039 & -1.679 & 0.09 \\ 
waiting time & 0.47 & 0.019 & 24.5 & $<0.001$ \\ 
waiting time$^2$ & -0.034 & 0.001 & -39.385 & $<0.001$ \\ \hline
NW vs NE & 0.266 & 0.062 & 4.317 & $<0.001$ \\  
Center vs NE & -0.197 & 0.055 & -3.58 & $<0.001$\\  
South + Is vs NE & -0.119 & 0.051 & -2.34 & 0.02 \\
Area 2 vs 1   & 0.04& 0.145 & 0.279 & 0.78 \\ 
Area 3 vs 1  & -0.104& 0.126 & -0.827 & 0.41\\ 
Area 4 vs 1  &  -0.114& 0.173 & -0.661 & 0.51 \\ 
Area 5 vs 1  & -0.393 & 0.111 & -3.558 & $<0.001$ \\ 
Area 6 vs 1  & -0.833 & 0.101 & -8.236 & $<0.001$ \\ 
Area 7 vs 1  & -0.322 & 0.122 & -2.645 & 0.008 \\ 
Area 8 vs 1   & -0.224 & 0.116 & -1.941 & 0.052 \\ 
Area 9 vs 1   & -0.012 & 0.113 & -0.11 & 0.91 \\
Area 10 vs 1   & -0.097 & 0.111 & -0.873 & 0.382 \\ 
Area 11 vs 1   & -0.118 & 0.112 & -1.051 & 0.293 \\ 
Area 12 vs 1  & -0.188 & 0.112 & -1.681 & 0.093 \\ 
 Area 13 vs 1  & -0.288 & 0.112 & -2.564 & 0.01 \\  
 Area 14 vs 1  & -0.131 & 0.133 & -0.98& 0.327 \\ 
 Year Entry 2005-10 vs $<2005$ & -1.444 & 0.052 & -27.763 & $<0.001$ \\ 
Year Entry $>=2010$  & -2.313 & 0.068 & -34.173 &$<0.001$ \\ 
\hline 
\end{tabular}
}
\end{table}

\subsection{Mobility}
The body of literature analyzing academic mobility is already extensive, see for example \citep{angervall2023academic} and \citep{yan2020analyzing}.
We reconstruct mobility in this data set. The 7.11\%  of faculties experiments one in life a change of University, 75\% of them change also the regions.  Generally, one of the motivations for moving is the opportunity for career advancement, our data confirm that mobility favour promotions. Introducing the dummy variable for changing region in the entire career in the logistic models illustrated in Tables \ref{tab:logistic1} and \ref{tab:logistic2}  we found a positive association between mobility and probability to be promoted. In particular, association with promotion from Associate to Full professors results 0.49 (95\% CI 0.38-0.59), and association with promotion from Grade E to Grade B results 
0.86 (95\% CI 0.67 1.05). In the last model, introducing mobility gender effects is not significative anymore. 
Change of University and region is more common in men than women,  68.2\% of mobility is within men. The probability to change region is negatively significant associate with being female with coefficient -0.18 (95\% CI -0.23--0.12).

\section{Conclusions}
\label{Conclusion}
In this study, we have explored the gender dynamics within Italian academia, particularly focusing on the promotion processes from Associate Professor to Full Professor. Through the analysis along the different macro-areas of research and political geographical region, we have shed light on the disparities and challenges faced by women in attaining higher academic positions.

Our findings reveal significant gender gaps persisting across various dimensions. Despite advancements over the past two decades, the proportion of women promoted to full professor remains notably lower than their male counterparts. Even within specific academic areas, such as STEM disciplines, where women are traditionally underrepresented, the disparity in promotion rates remains stark.

Through logistic regression models incorporating macro-area of research, region, year, waiting time, we have quantified these gender disparities and explored potential interactions. The results underscore the need for targeted interventions and policies to address systemic barriers faced by women in academia.

Despite regional differences across Italy, our study underscores a consistent pattern of gender disparity in academic promotion across all regions, from the North to the South. This uniformity highlights the pervasive nature of the challenges faced by women in advancing their academic careers, irrespective of geographic location. It emphasizes the need for comprehensive and inclusive strategies to address gender inequality in academia that transcend regional boundaries and address systemic barriers at a national level.

As we move forward, it is imperative to continue monitoring and addressing gender disparities in academic promotions. Initiatives aimed at promoting diversity, equity, and inclusion are crucial for fostering a more equitable and supportive academic environment. Additionally, further research employing longitudinal and comprehensive datasets, along with advanced statistical techniques like Cox analysis to handle censoring, will provide deeper insights into the underlying mechanisms driving gender disparities in academic advancement.

Ultimately, achieving gender parity in academic leadership positions is not only a matter of equity but also essential for fostering innovation, excellence, and diversity in the academic landscape. It is our collective responsibility to strive towards creating an inclusive and equitable academic environment where talent and merit are recognized irrespective of gender.
 
{\tiny
\begin{longtable}[c]{r l l l }
\caption{Name of the University, Kind (S=National, P=Private, T=Online), Geographical Region. For online universities, the geographical region is left blank.\label{region}}\\
 
\hline
 & Name & Kind & Region  \\ 
 \hline
 \endfirsthead
\hline
 & Name & Kind & Region  \\ 
 \hline
 \endhead
 \hline
 \endfoot
1 & BARI & S & PUGLIA  \\ 
  2 & BASILICATA & S & BASILICATA  \\ 
  3 & BERGAMO & S & LOMBARDIA  \\ 
  4 & Bocconi MILANO & P & LOMBARDIA  \\ 
  5 & BOLOGNA & S & EMILIA ROMAGNA  \\ 
  6 & BRESCIA & S & LOMBARDIA  \\ 
  7 & Ca' Foscari VENEZIA & S & VENETO   \\ 
  8 & CAGLIARI & S & SARDEGNA  \\ 
  9 & CAMERINO & S & MARCHE  \\ 
  10 & CAMPANIA - L. VANVITELLI & S & CAMPANIA  \\ 
  11 & CASSINO e LAZIO MERIDIONALE & S & LAZIO  \\ 
  12 & CATANIA & S & SICILIA  \\ 
  13 & CATANZARO & S & CALABRIA  \\ 
  14 & Cattolica del Sacro Cuore & P & LOMBARDIA  \\ 
  15 & CHIETI-PESCARA & S & ABRUZZO  \\ 
  16 & della CALABRIA & S & CALABRIA  \\ 
  17 & EUROPEA di ROMA & P & LAZIO  \\ 
  18 & FERRARA & S & EMILIA ROMAGNA   \\ 
  19 & FIRENZE & S & TOSCANA   \\ 
  20 & FOGGIA & S & PUGLIA  \\ 
  21 & GENOVA & S & LIGURIA  \\ 
  22 & Gran Sasso Science Institute & S & ABRUZZO  \\ 
  23 & HUMANITAS University & P & LOMBARDIA  \\ 
  24 & I.U.S.S. - PAVIA & S & LOMBARDIA  \\ 
  25 & INSUBRIA & S & LOMBARDIA  \\ 
  26 & IULM - MILANO & P & LOMBARDIA  \\ 
  27 & L'AQUILA & S & ABRUZZO  \\ 
  28 & L'Orientale di NAPOLI & S & CAMPANIA  \\ 
  29 & Libera Univ. Maria SS.Ass.-LUMSA-ROMA & P & LAZIO  \\ 
  30 & Libera Universita di BOLZANO & P & TRENTINO A. A.  \\ 
  31 & LINK CAMPUS & T &  \\ 
  32 & LIUC - CASTELLANZA & P & LOMBARDIA  \\ 
  33 & Luiss Guido Carli & P & LAZIO  \\ 
  34 & LUM Giuseppe Degennaro & P & PUGLIA  \\ 
  35 & MACERATA & S & MARCHE  \\ 
  36 & Mediterranea di REGGIO CALABRIA & S & CALABRIA  \\ 
  37 & MESSINA & S & SICILIA  \\ 
  38 & MILANO & S & LOMBARDIA  \\ 
  39 & MILANO-BICOCCA & S & LOMBARDIA   \\ 
  40 & MODENA e REGGIO EMILIA & S & EMILIA ROMAGNA  \\ 
  41 & MOLISE & S & MOLISE  \\ 
  42 & Napoli Federico II & S & CAMPANIA  \\ 
  43 & PADOVA & S & VENETO   \\ 
  44 & PALERMO & S & SICILIA  \\ 
  45 & PARMA & S & EMILIA ROMAGNA  \\ 
  46 & Parthenope di NAPOLI & S & CAMPANIA  \\ 
  47 & PAVIA & S & LOMBARDIA  \\ 
  48 & PERUGIA & S & UMBRIA  \\ 
  49 & PIEMONTE ORIENTALE & S & PIEMONTE  \\ 
  50 & PISA & S & TOSCANA  \\ 
  51 & Politecnica delle MARCHE & S & MARCHE  \\ 
  52 & Politecnico di BARI & S & PUGLIA  \\ 
  53 & Politecnico di MILANO & S & LOMBARDIA  \\ 
  54 & Politecnico di TORINO & S & PIEMONTE  \\ 
  55 & ROMA Foro Italico & S & LAZIO  \\ 
  56 & ROMA La Sapienza & S & LAZIO   \\ 
  57 & ROMA Tor Vergata & S & LAZIO  \\ 
  58 & ROMA TRE & S & LAZIO  \\ 
  59 & S. Raffaele MILANO & P & LOMBARDIA  \\ 
  60 & SALENTO & S & PUGLIA  \\ 
  61 & SALERNO & S & CAMPANIA  \\ 
  62 & SANNIO di BENEVENTO & S & CAMPANIA  \\ 
  63 & SASSARI & S & SARDEGNA  \\ 
  64 & SCIENZE GASTRONOMICHE & P & PIEMONTE  \\ 
  65 & Scuola IMT Alti Studi LUCCA & S & TOSCANA  \\ 
  66 & Scuola Normale Superiore di PISA & S & TOSCANA  \\ 
  67 & Scuola Superiore Sant'Anna & S & TOSCANA   \\ 
  68 & SIENA & S & TOSCANA  \\ 
  69 & SISSA - TRIESTE & S & FRIULI V. G.  \\ 
  70 & Stranieri di PERUGIA & S & UMBRIA  \\ 
  71 & Stranieri di SIENA & S & TOSCANA  \\ 
  72 & Stranieri REGGIO CALABRIA & S & CALABRIA  \\ 
  73 & Suor Orsola Benincasa - NAPOLI & P & CAMPANIA  \\ 
  74 & TERAMO & S & ABRUZZO  \\ 
  75 & TORINO & S & PIEMONTE  \\ 
  76 & TRENTO & S & TRENTINO A. A.  \\ 
  77 & TRIESTE & S & FRIULI V. G.  \\ 
  78 & TUSCIA & S & LAZIO  \\ 
  79 & UDINE & S & FRIULI V. G.  \\ 
  80 & UKE - Universita Kore di ENNA & P & SICILIA  \\ 
  81 & UniCamillus - Saint Camillus International & P & LAZIO  \\ 
  82 & UNICUSANO - Telematica Roma & T   \\ 
  83 & Univ. Campus Bio-Medico di ROMA & P & LAZIO  \\ 
  84 & Univ. Studi GUGLIELMO MARCONI-T. & T &    \\ 
  85 & Univ. Studi Intern. di ROMA (UNINT) & P & LAZIO  \\ 
  86 & Univ. Telematica E-CAMPUS & T &    \\ 
  87 & Univ. Telem. GIUSTINO FORTUNATO & T   &  \\ 
  88 & Univ. Telem. Intern. UNINETTUNO & T   &  \\ 
  89 & Univ. Telematica IUL & T &    \\ 
  90 & Univ. Telematica LEONARDO da VINCI & T   &  \\ 
  91 & Univ. Telematica PEGASO & T &    \\ 
  92 & Univ. Telematica San Raffaele Roma & T &    \\ 
  93 & Univ. Telematica UNITELMA SAPIENZA & T &   \\ 
  94 & Univ. Telem. Universitas MERCATORUM & T &    \\ 
  95 & Universita IUAV di VENEZIA & S & VENETO  \\ 
  96 & Urbino Carlo Bo & S & MARCHE  \\ 
  97 & VALLE D'AOSTA & S & VALLE D'AOSTA   \\ 
  98 & VERONA & S & VENETO  \\    
\end{longtable}
}

\bibliography{gender_leap.bib}

\end{document}